\documentclass[manuscript]{acmart}

\usepackage{multirow}%
\usepackage{makecell}%
\AtBeginDocument{%
  }

\setcopyright{acmlicensed}
\copyrightyear{2018}
\acmYear{2018}
\acmDOI{XXXXXXX.XXXXXXX}
\acmConference[Conference acronym 'XX]{Make sure to enter the correct
  conference title from your rights confirmation email}{June 03--05,
  2018}{Woodstock, NY}
\acmISBN{978-1-4503-XXXX-X/2018/06}

\usepackage{mdframed}%
\newmdenv[
  linecolor=black,
  linewidth=0.5pt,
  leftmargin=0pt,
  rightmargin=0pt,
  skipabove=0pt,
  skipbelow=0pt,
  innerleftmargin=2pt,
  innerrightmargin=2pt,
  innertopmargin=2pt,
  innerbottommargin=2pt,
]{promptbox}

\newcommand{\prompt}[2]{%
  \begin{center}
    \textbf{#1}
  \end{center}
  \vspace{1pt}
  {\setlength{\parskip}{0pt}%
  \begin{promptbox}
    \ttfamily \footnotesize
    \setstretch{1}
    #2
  \end{promptbox}}
}

\begin{document}
\renewcommand\footnotetextcopyrightpermission[1]{}%
\settopmatter{printacmref=false} %remove ACM reference format
% \title{The Name of the Title Is Hope}
\title{Generalized Multi-agent Social Simulation Framework}

\author{Gang Li}
\authornote{Equal contribution}
\email{}
\orcid{}
\affiliation{%
  \institution{Zhejiang University}
  \city{Hangzhou}
  \state{Zhejiang}
  \country{China}
}

\author{Jie Lin}
\authornotemark[1]
\email{}
\orcid{}
\affiliation{%
  \institution{Zhejiang University}
  \city{Hangzhou}
  \state{Zhejiang}
  \country{China}
}

\author{Yining Tang}
\authornotemark[1]
\email{}
\orcid{}
\affiliation{%
  \institution{Zhejiang University}
  \city{Hangzhou}
  \state{Zhejiang}
  \country{China}
}

\author{Ziteng Wang}
\authornotemark[1]
\email{}
\orcid{}
\affiliation{%
  \institution{Zhejiang University}
  \city{Hangzhou}
  \state{Zhejiang}
  \country{China}
}

\author{Yirui Huang}
\email{}
\orcid{}
\affiliation{%
  \institution{Zhejiang University}
  \city{Hangzhou}
  \state{Zhejiang}
  \country{China}
}

\author{Junyu Zhang}
\email{}
\orcid{}
\affiliation{%
  \institution{Zhejiang University}
  \city{Hangzhou}
  \state{Zhejiang}
  \country{China}
}

\author{Shuang Luo}
\email{}
\orcid{}
\affiliation{%
  \institution{Zhejiang University}
  \city{Hangzhou}
  \state{Zhejiang}
  \country{China}
}

\author{Yike Guo}
\email{}
\orcid{}
\affiliation{%
  \institution{Hong Kong University of Science and Technology}
  \city{Hong Kong}
  \country{China}
}

\author{Chao Wu}
\authornote{Corresponding author}
\email{}
\orcid{}

\affiliation{%
  \institution{Zhejiang University}
  \city{Hangzhou}
  \state{Zhejiang}
  \country{China}
}

% Compact numbered affiliations in the manuscript author block.
% The individual author/affiliation fields above remain available to ACM metadata.
\makeatletter
\gdef\@currentauthors{}
\gdef\@currentaffiliations{}
\def\@mkauthors@i{%
  \global\setbox\mktitle@bx=\vbox{%
    \noindent\unvbox\mktitle@bx\par\medskip
    \noindent{\@authorfont
      \mbox{GANG LI\textsuperscript{1,*}},\enspace
      \mbox{JIE LIN\textsuperscript{1,*}},\enspace
      \mbox{YINING TANG\textsuperscript{1,*}},\enspace
      \mbox{ZITENG WANG\textsuperscript{1,*}},\enspace
      \mbox{YIRUI HUANG\textsuperscript{1}}\par\noindent
      \mbox{JUNYU ZHANG\textsuperscript{1}},\enspace
      \mbox{SHUANG LUO\textsuperscript{1}},\enspace
      \mbox{YIKE GUO\textsuperscript{2}},\enspace
      \mbox{CHAO WU\textsuperscript{1,\textdagger}}}\par\smallskip
    \noindent{\@affiliationfont
      \textsuperscript{1} Zhejiang University, Hangzhou, China\par
      \textsuperscript{2} Hong Kong University of Science and Technology, Hong Kong, China}\par\medskip
  }%
}
\makeatother
\authorsaddresses{}

\renewcommand{\shortauthors}{Li et al.}

% \begin{abstract}
%   A clear and well-documented \LaTeX\ document is presented as an
%   article formatted for publication by ACM in a conference proceedings
%   or journal publication. Based on the ``acmart'' document class, this
%   article presents and explains many of the common variations, as well
%   as many of the formatting elements an author may use in the
%   preparation of the documentation of their work.
% \end{abstract}
\begin{abstract}
Multi-agent social interaction has clearly benefited from Large Language Models.
However, current simulation systems still face challenges such as difficulties in scaling to diverse scenarios and poor reusability due to a lack of modular design.
To address these issues, we designed and developed a modular, object-oriented framework that organically integrates various base classes through a hierarchical structure, harvesting scalability and reusability.
We inherited the framework to realize common derived classes.
Additionally, a memory summarization mechanism is proposed to filter and distill relevant information from raw memory data, prioritizing contextually salient events and interactions.
By selecting and combining some necessary derived classes, we customized a specific simulated environment.
Utilizing this simulated environment, we successfully simulated human interactions on social media, replicating real-world online social behaviors.
The source code for the project will be released and evolve.
\end{abstract}




\maketitle

% \section{Introduction}
% ACM's consolidated article template, introduced in 2017, provides a
% consistent \LaTeX\ style for use across ACM publications, and
% incorporates accessibility and metadata-extraction functionality
% necessary for future Digital Library endeavors. Numerous ACM and
% SIG-specific \LaTeX\ templates have been examined, and their unique
% features incorporated into this single new template.

% If you are new to publishing with ACM, this document is a valuable
% guide to the process of preparing your work for publication. If you
% have published with ACM before, this document provides insight and
% instruction into more recent changes to the article template.

% The ``\verb|acmart|'' document class can be used to prepare articles
% for any ACM publication --- conference or journal, and for any stage
% of publication, from review to final ``camera-ready'' copy, to the
% author's own version, with {\itshape very} few changes to the source.

\section{Introduction}

In the field of social simulation~\cite{tesfatsion2003agent,radzicki2008origin,adamatzky2010game,gustafsson2010consistent,zheng2020ai}, researchers aim to create highly credible artificial environments that are closely aligned with human society (e.g., life simulation games such as \textit{The Sims}, complex cognitive models~\cite{kotseruba202040,samsonovich2010toward} and virtual environments~\cite{li2023metaagents}).
Within these environments, agents are designed to realistically simulate human behavior.

In recent years, significant progress has been made in the development of generative agents~\cite{kotseruba202040,park2023generative}, which refer to agents capable of generating and simulating human behavior, language, and thought process.
Generative agents have expanded the boundaries of simulating human behavior~\cite{guo2024large} across a diverse range of activities.
These activities encompass not only everyday tasks, for example getting up and making breakfast, but also more complex behavior like reflecting on past experiences, forming opinions and engaging in conversations.
These advancements demonstrate the potential of generative agents in simulating complex human behavior.

Despite these new advances in virtual worlds and social simulations, challenges remain.
Current frameworks suffer from low scalability and reusability and a lack of components similar to those of human society.
In addition, memory systems are burdened with redundant information, leading to a low utilization efficiency in leveraging stored content.

\begin{figure}[!htb]
    \centering
    \includegraphics[width=1.0\linewidth]{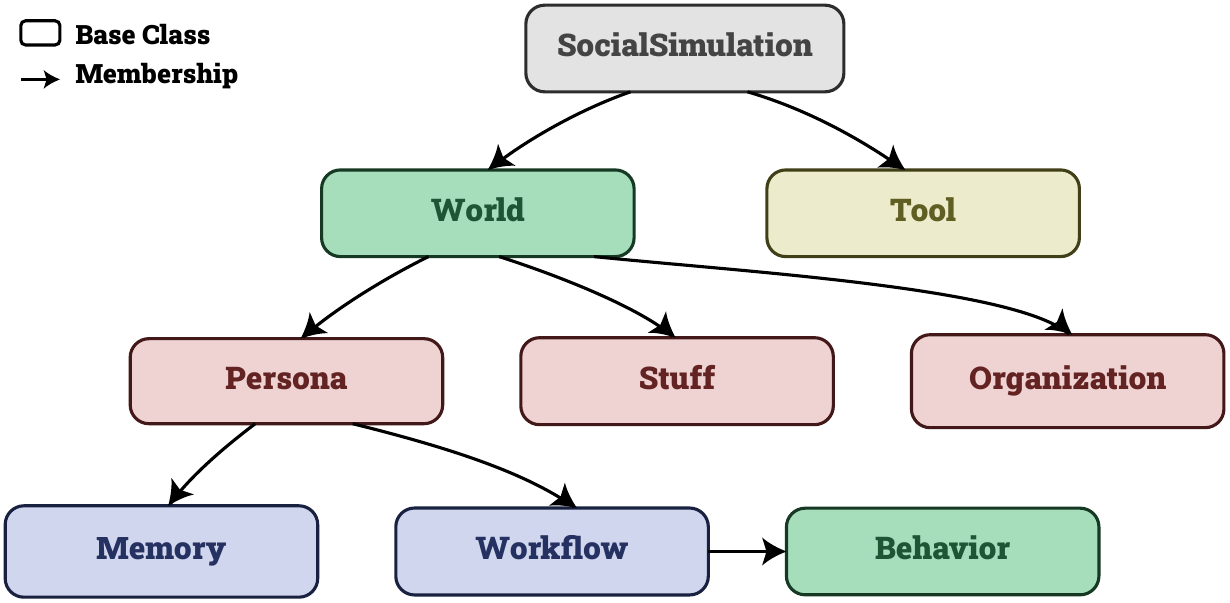}
    \caption{The hierarchical structure of base classes.}
    \Description{}
    \label{fig:framework}
\end{figure}

To this end, we propose an innovative solution.
As illustrated in Figure~\ref{fig:framework}, we design a modular, object-oriented framework where the components are organically integrated through a hierarchical structure.
Each component is a generic base class corresponding to the various components of human society, and specific simulated environments are created by inheriting these base classes.
We realize common derived classes which others can simply reuse.
Moreover, we propose a memory summarization mechanism that filters and distills relevant information from raw memory data, reducing redundancy.

We selected some off-the-shelf derived classes to customize a specific simulated environment.
The simulation scenario was designed to represent online activities in the real world, where agents engage in discussions on social issues and interact within the virtual community.
We conducted a series of experiments to validate the effectiveness of our customized environment.

In summary, the main contributions of this paper are as follows:

\textbf{Modular, Object-Oriented System Design.}
The framework has a modular and object-oriented design, where each component serves as a general base class. 
By inheriting these base classes, the framework can be adapted to different simulated environments, offering a high degree of scalability and reusability.

\textbf{Memory Summarization Mechanism.}
This mechanism reduces information overload, prioritizes contextually salient events and interactions, and enables agents to exhibit more adaptive and cognitively plausible behaviors by focusing on meaningful historical data.

\section{Related Work}

\subsection{Complex Social Simulation}

Complex social simulation is a rapidly evolving field that uses agent-based modeling and system dynamics to study social, natural, and information sciences~\cite{oren1994artificial}. An evolutionary constructive approach can help understand the intricate dynamics of complex social systems~\cite{hashimoto2008evolutionary}. A complex adaptive framework has been proposed to address the interaction between macro social constraints and individual agency in human development~\cite{abdollahian2013human}. Complexity and emergence in communication networks can lead to consequences ranging from increased costs to network shutdowns~\cite{niazi2013complex}. Agent-based models have been used to predict and control epidemics in large-scale urban environments~\cite{zhang2016large}. Incorporating social sciences in simulation studies enhances the realism of complex urban environments~\cite{chai2016using}. Risk diffusion models based on the SIR model have been developed to analyze social stability risks in large complex engineering projects~\cite{he2021social}. Non-identifiability challenges in social simulation have led to proposals for high-fidelity calibration objectives in financial market simulations~\cite{wang2024alleviating}.

\subsection{LLM-based Multi-Agent Framework}

In recent years, advancements in artificial intelligence have led to new frameworks for multi-agent interactions based on Large Language Models (LLMs). \citet{jinxin2023cgmi} introduces CGMI with a tree-structured persona model to maintain role consistency across dialogues. The Artificial General Agent~\cite{yu2024affordable} model improves cost-efficiency through its Lifestyle Policy and Social Memory modules. Despite its benefits, AGA struggles with behavior diversity in static settings, often resulting in predictable interactions.
Further developments include AutoGen~\cite{wu2024autogen}, a multi-agent dialogue framework allowing LLM-based application development, and the Internet of Agents (IoA)~\cite{chen2024internet}, which offers a platform for agent collaboration with innovative communication and teaming mechanisms. Additionally, METAAGENTS~\cite{li2023metaagents} presents agents with collaborative and human-like reasoning abilities for complex tasks.
These frameworks can be applied in a variety of ways including solving problems~\cite{qian2024chatdev,chan2023chateval,yu2023co}, simulating the real world~\cite{liutraining,mou2024unveiling,hongmetagpt}.

\subsection{LLM-based Multi-Agent Simulation}

Recent advances in Large Language Models (LLMs) have opened up new possibilities for creating more complex and realistic multi-agent simulations. These advances have broad implications, covering a variety of areas including social simulation~\cite{park2022social,gao2023s,li2023quantifying}, game simulation~\cite{xu2023exploring,xu2023language,light2023avalonbench,wang2023avalon,mukobi2023welfare}, psychological simulation~\cite{aher2023using,zhang2023exploring}, economy simulation~\cite{weissrethinking,li2023tradinggpt,zhao2023competeai}, recommender systems~\cite{zhang2024generative,zhang2024agentcf}, and policy making~\cite{xiao2023simulating,hua2023war}. \citet{park2023generative}
introduced generative agents that simulate human behavior, laying the groundwork for interactive environments mimicking daily activities and decision-making.
LLM-powered agents in economic simulations~\cite{li2023large} demonstrated human-like decision-making, producing more plausible macroeconomic phenomena than traditional models.
The RecAgent paradigm~\cite{wang2023large} used LLMs to simulate user interactions with recommender systems, highlighting the importance of multi-module design for consistent behavior.
LLM-based agents were applied to simulate opinion evolution in social networks~\cite{chuang2023simulating}, incorporating factors like confirmation bias and memory updating.
The SOTOPIA environment~\cite{zhousotopia} advanced social intelligence evaluation by simulating complex social interactions, revealing differences between artificial and human social performance.

% \section{Modifications}

% Modifying the template --- including but not limited to: adjusting
% margins, typeface sizes, line spacing, paragraph and list definitions,
% and the use of the \verb|\vspace| command to manually adjust the
% vertical spacing between elements of your work --- is not allowed.

% {\bfseries Your document will be returned to you for revision if
%   modifications are discovered.}

\section{Method}

\subsection{Basic Framework Explanation}

\begin{figure*}[!htb]
    \includegraphics[width=1.0\textwidth]{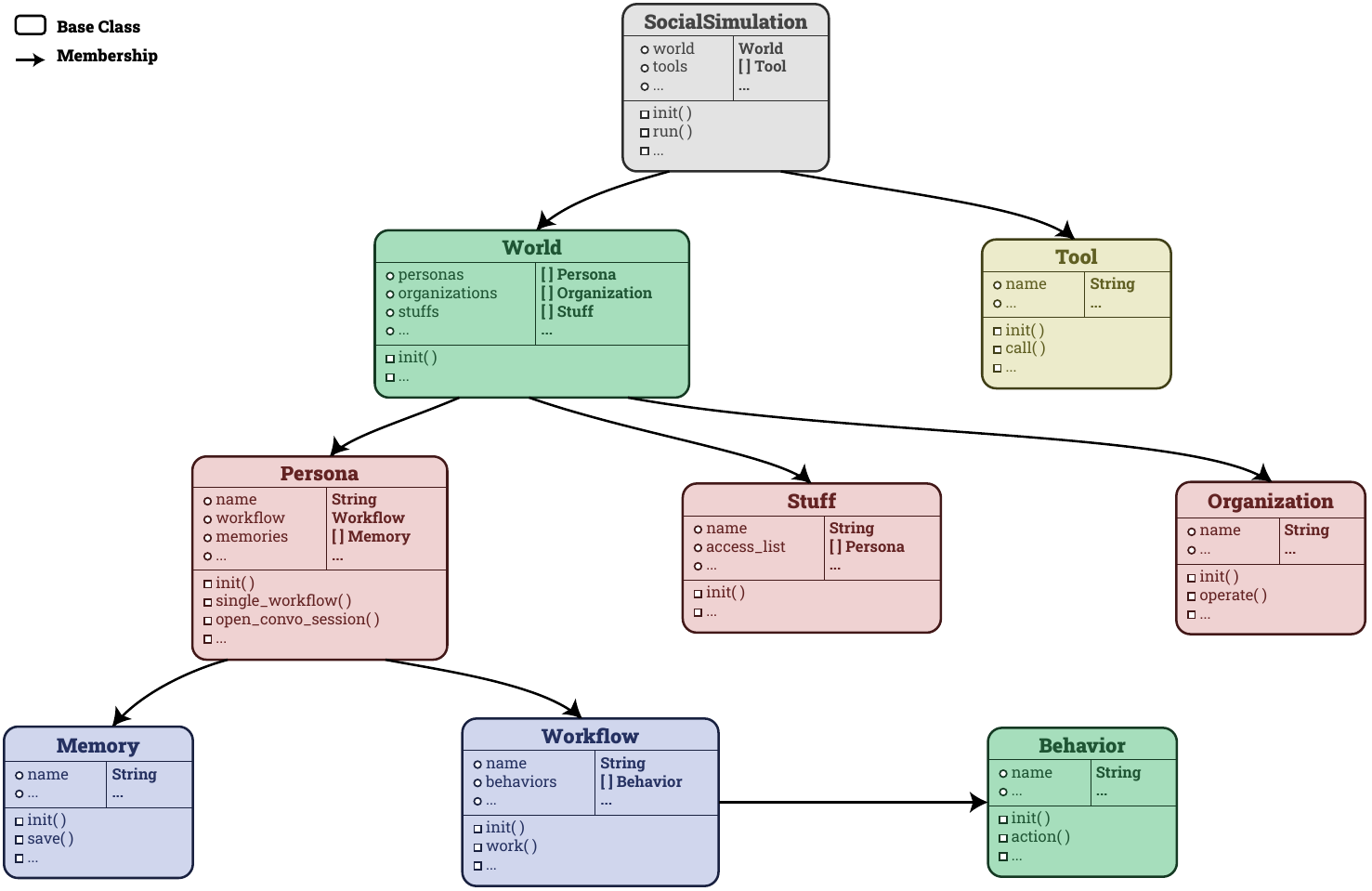}
    \caption{The hierarchical structure of the proposed framework.}
    \Description{}
    \label{fig:general_uml}
\end{figure*}

To enhance scalability and reusability, we introduced a novel architecture with a modular, object-oriented design.
As shown in Figure~\ref{fig:general_uml}, this architecture is composed of various core components and each component is defined as a generic base class, and specific derived classes are created by inheriting these base classes to meet distinct requirements, as detailed below.

\textbf{Social Simulation.} It is responsible for managing the components and controlling the entire simulation process.

\textbf{World.} The module can be customized to various worlds depending on specific needs, for example offline physical world and online virtual world.

\textbf{Tool.} Through inheritance, this module enables the creation of various tools, such as search engines. 
It is designed with the Singleton Pattern~\cite{gamma1994design} to ensure efficient use and management, avoiding resource waste and complexity from multiple tools while maintaining consistent tool behavior.

\textbf{Persona.} The persona is the agent which is autonomous and capable of mimicking human behavior, interacting with other agents and the environment.

\textbf{Stuff.} This refers to all available resources in the world, including both abstract entities, such as events and policies, and concrete entities, for example non-player characters (NPCs).

\textbf{Organization.} We introduced this module to simulate organization, for example companies, government bodies, and civic groups, providing realistic social dynamics to enhance the realism.

\textbf{Memory.} The memory module stores an agent's initial state, environmental perceptions, and internal reflections, guiding its functions. 
Different memory modules can be created as needed to store various types of information, such as short-term memory, long-term memory, and spatial memory.

\textbf{Workflow.} This module guides agents through a structured sequence of behaviors to accomplish tasks efficiently and coherently.
Using the Strategy Pattern~\cite{gamma1994design}, different behaviors can be combined to form different workflows.

\textbf{Behavior.} Each one represents a specific human behavior, such as perception and decision-making, simulating the specific operations a person would perform when completing tasks.

Furthermore, Workflow and Memory adopt the Visitor Pattern~\cite{gamma1994design}, allowing the workflow module to flexibly access, operate, or combine multiple types of memory without altering their data structure.

\begin{figure*}
\Description{}
\prompt{Memory Summary}{

\textbf{\#\# Key Question Prompt}

System Instruction:

You are \{persona\_name\} and your profile is \{profile\}. Based on the target character's question and profile, output the key question you need to know to answer the question.Please output the results in the following format.

-

[Output format]

[key\_question1, key\_question2, ...]

-

User Prompt:

\{source\_persona\_name\}’s profile is:

\{source\_persona\_profile\}

-

The question from \{source\_persona\_name\} is:

\{raw\_question\}

-

\textbf{\#\# Summary Prompt}

System Instruction:

Answer this question according to the following content.

-

User Prompt:

The content is:

\{retrieved\_content $R$\}

-

The question is:

\{key\_quesiton\}

}
\caption{The prompt template for memory summary.} % Caption of the figure
\label{fig: memory_summary} % Associated with autoref, set cross-reference and display "Figure x.x"
\end{figure*}

\subsection{Memory Summarization Mechanism}

In the process of agents calling memory for decision-making or dialogue, the large language model first generates multiple retrieval keywords and performs query operations in the memory in sequence. For each keyword, we select the top-k most relevant results as the final retrieval results $R$.

On this basis, to improve agents' understanding of memory content and enable them to generate decisions and dialogues more suitable for the scenario, we propose a summarization mechanism for further generalization and in-depth analysis of the final retrieval results $R$. Specifically, this mechanism first calls the large language model to determine $P$ key information points according to current task requirements, and then allows the large language model to extract corresponding relevant information $C_i$ from the final retrieval results $R$ for each key information point. Finally, the refined information $[C_1, C_2, ..., C_P]$ is provided to agents to assist their decision-making and dialogue generation. The prompt template for summarization is shown in Figure~\ref{fig: memory_summary}.

% \section{Typefaces}

% The ``\verb|acmart|'' document class requires the use of the
% ``Libertine'' typeface family. Your \TeX\ installation should include
% this set of packages. Please do not substitute other typefaces. The
% ``\verb|lmodern|'' and ``\verb|ltimes|'' packages should not be used,
% as they will override the built-in typeface families.

\section{Experiment}

By inheriting the proposed framework, we realize common derived classes, which can be found in Appendix~\ref{app:allsubclasses}.
We conduct experimental evaluations from two dimensions: controlled evaluation and end-to-end evaluation. Specifically: (1) At the level of controlled evaluation, the proposed framework is systematically verified through benchmark to quantitatively assess the functional accuracy; (2) At the level of end-to-end evaluation, multi-turn conversation experiments are adopted to observe the behavioral evolution of agent group in complex social situations and detect whether human-like behavioral patterns emerge.

\subsection{Controlled Evaluation}

\begin{figure*}
\Description{}
\prompt{GPT Evaluation}{

System Instruction:

You are a helpful assistant, that ranks models by the quality of their responses.

-

User Prompt:

I want you to create a leaderboard for different large-language models. For this purpose, I will provide a summary of a complete script along with its related questions and the respective responses from the two models. Ensure that your ranking is impartial concerning the position of the models. The evaluation should be based on the following criteria:

1. Relevance to the question.

2. Reflection of the role's characteristics.

3. Overall quality of the response, including fluency, coherence, and language expression.

-

Here is the question:

\{\{"question": \{question\}\}\}

Here is the role they play:

\{\{"role": \{role\_name\}\}\}

Here is the summary of the script:

\{\{"summary": \{summary\}\}\}

Here are the responses of two models: \{list\_model\_answer\_dict\}

-

Give a reason to score each model output and return the model name with a higher score. Output a Python dictionary formatted as follows:

\{\{

    "model\_1": <rank-reason>,
    
    "model\_2": <rank-reason>,
    
    "winner": 1(represents model\_1) or 2(represents model\_2).
    
\}\}

Your response must be a valid Python dictionary and should contain nothing else, as it will be directly executed in Python.

}
\caption{The prompt template for GPT evaluation.} % Caption of the figure
\label{fig: roleagentbench_evaluate_prompt} % Associated with autoref, setting cross-reference and displaying "Figure x.x"
\end{figure*}

For controlled evaluation, experiments are conducted on the RoleAgentBench~\cite{liu2024roleagent} dataset. RoleAgentBench is a systematic and comprehensive evaluation benchmark designed to assess the generation ability and interaction performance of agents. The dataset covers 20 original English scripts and 5 original Chinese scripts, containing a total of 128 characters (100 English characters and 28 Chinese characters), spanning various scenarios such as movies and TV dramas. The evaluation system of RoleAgentBench consists of two parts: agent simulation and memory system. The agent simulation part includes three core evaluation tasks: Self-Knowledge, Reaction, and General Response. Among them, Self-Knowledge and Reaction are evaluated using Accuracy (ACC), and General Response is evaluated using Win Rate (WR). The memory system part mainly uses WR as an indicator to measure the quality of the memory system. All evaluation questions are based on the original scripts, generated by GPT-4 and manually calibrated to ensure their high consistency with the character settings, thereby providing a multi-dimensional and multi-lingual standardized testing benchmark for agents. The evaluation metrics for this test mainly include two categories: ACC and WR. The calculation of ACC is shown in Equation \ref{eq: acc}.
\begin{align}
    \label{eq: acc}
    ACC = \frac{\sum_i^N I(y_i == \hat{y_i}) }{N_{all}}
\end{align}
where $N$ is the total number of samples in the test set, $y_i$ is the predicted result of the $i$-th test sample, $\hat{y_i}$ is the true label of the $i$-th test sample, and $I(\cdot)$ is an indicator function, which takes a value of 1 when $y_i = \hat{y_i}$, and 0 otherwise. For WR, its calculation is shown in Equation \ref{eq: agent_win_rate}.
\begin{align}
\label{eq: agent_win_rate}
WR = \frac{\sum_i^N I(Rank(y_i) == 1)}{N_{all}}
\end{align}
where $N$ is the total number of samples in the test set, $Rank(\cdot)$ is a ranking function used to rank the model's predicted answers against the reference answers, and $I(\cdot)$ is an indicator function, which takes a value of 1 if the rank is first, and 0 otherwise. The ranking process employs prompt engineering to enable GPT-4 for ranking, and its prompt template is shown in Figure~\ref{fig: roleagentbench_evaluate_prompt}.

\subsubsection{Result}

\begin{table}[htbp]
	\centering
    \renewcommand\arraystretch{1.5}
	\caption{Experimental results on RoleAgentBench}\label{tab: multi_agent_performance_res}
    \setlength{\tabcolsep}{5pt}
    \begin{tabular}{ccccc}
    \hline
    \multirow{2}{*}{\textbf{Method}} & \multicolumn{3}{c}{\textbf{Agent Simulation}} & \multirow{2}{*}{\textbf{Memory System}} \\  % For the aesthetics of table typesetting, it is recommended to bold the table headers.
    \cmidrule{2-4}
     & \textbf{\makecell[c]{Self-\\Knowledge}} & \textbf{Reaction} & \textbf{\makecell[c]{General\\Response}} & \\
    \hline
    Generative Agents~\cite{park2023generative} & 68.90 & 40.70 & 31.80 & 38.90 \\
    RoleAgent~\cite{liu2024roleagent} & 85.60 & 56.90 & 31.90 & 41.30 \\
    \textbf{Ours} & \textbf{91.35} & \textbf{60.00} & \textbf{32.64} & \textbf{43.90} \\
    \hline
    \end{tabular}%
\end{table}

To verify the effectiveness of the proposed framework in complex multi-agent scenarios, this paper conducts multi-dimensional comparative experiments based on the RoleAgentBench benchmark. It can be seen from Table~\ref{tab: multi_agent_performance_res} that compared with Generative Agents~\cite{park2023generative} and RoleAgent~\cite{liu2024roleagent}, the proposed framework has shown improvements in the four dimensions of Self-Knowledge, Reaction, General Response, and Memory System.

\subsubsection{Ablation Study}

\begin{table}[htbp]
	\centering
    \renewcommand\arraystretch{1.5}
	\caption{Ablation results on RoleAgentBench}\label{tab: multi_agent_ablation}
    \setlength{\tabcolsep}{5pt}%1pt
    \begin{tabular}{ccccc}
    \hline
    \multirow{2}{*}{\textbf{Method}} & \multicolumn{3}{c}{\textbf{Agent Simulation}} & \multirow{2}{*}{\textbf{Memory System}} \\  % For the aesthetics of table typesetting, it is recommended to bold the table headers.
    \cmidrule{2-4}
     & \textbf{\makecell[c]{Self-\\Knowledge}} & \textbf{Reaction} & \textbf{\makecell[c]{General\\Response}} &  \\
    \hline
    Ours - Without summary & 78.85 & 57.27 & 30.91 & 25.60 \\
    \textbf{Ours} & \textbf{91.35} & \textbf{60.00} & \textbf{32.64} & \textbf{43.90} \\
    \hline
    \end{tabular}%
\end{table}

\begin{table*}[h]
\centering
\scriptsize
\caption{Profiles of Individuals}
\label{tab:profile}
\begin{tabular}{>{\raggedright\arraybackslash}p{0.11\textwidth}|>{\raggedright\arraybackslash}p{0.84\textwidth}}
\hline
\textbf{\makecell[c]{Name}} & \textbf{\makecell[c]{Profile}} \\
\hline
\makecell[c]{Isabella Rodriguez}&\multicolumn{1}{m{0.84\textwidth}}{Isabella Rodriguez, 34, friendly, outgoing, hospitable. Isabella Rodriguez is the owner of Hobbs Cafe who loves to make people feel welcome. She constantly seeks ways to turn the cafe into a relaxing and enjoyable place for visitors. Isabella Rodriguez is planning on having a Valentine's Day party at Hobbs Cafe with her customers on February 14th, 2023 at 5pm. She is gathering party material, and is telling everyone to join the party at Hobbs Cafe on February 14th, 2023, from 5pm to 7pm.}\\
\hline
\makecell[c]{Maria Lopez} & \multicolumn{1}{m{0.84\textwidth}}{Maria Lopez, 21, energetic, enthusiastic, inquisitive. Maria Lopez is a student at Oak Hill College and a part-time Twitch game streamer, loves connecting with people and exploring new ideas. Maria Lopez is working on her physics degree and streaming games on Twitch to make some extra money. She visits Hobbs Cafe for studying and eating just about everyday.} \\
\hline
\makecell[c]{Klaus Mueller} & \multicolumn{1}{m{0.84\textwidth}}{Klaus Mueller, 20, kind, inquisitive, passionate. Klaus Mueller is a student at Oak Hill College studying sociology. He is passionate about social justice and enjoys exploring diverse perspectives. Klaus Mueller is writing a research paper on the effects of gentrification in low-income communities.} \\
\hline
\end{tabular}
\label{tab:profiles}
\end{table*}

We conduct an ablation experiment on the memory summarization mechanism using the RoleAgentBench dataset. The experimental results in Table~\ref{tab: multi_agent_ablation} indicate that the memory summarization mechanism has an improving effect on the simulation performance of the proposed framework.

\subsection{End-to-end Evaluation}

For end-to-end evaluation, we design multiple experiments covering various scenarios, from basic single-event discussions to interwoven multi-event discussions, committee policy releases, and applications of search engine results. Each experiment involves three agents, and their behaviors are observed through three rounds of discussions. The basic details of the agents are presented in Table~\ref{tab:profile}. Their personalized characteristics and backgrounds ensure the richness and diversity of the discussions, thereby more truthfully reflecting the collision of diverse perspectives caused by individual differences in social contexts. By simulating a series of real internet communications, we obtain relevant discussion contexts that conform to the characteristics of actual online interactions, verifying the effectiveness of the proposed framework.

\begin{figure*}[!htb]
  \includegraphics[width=\textwidth]{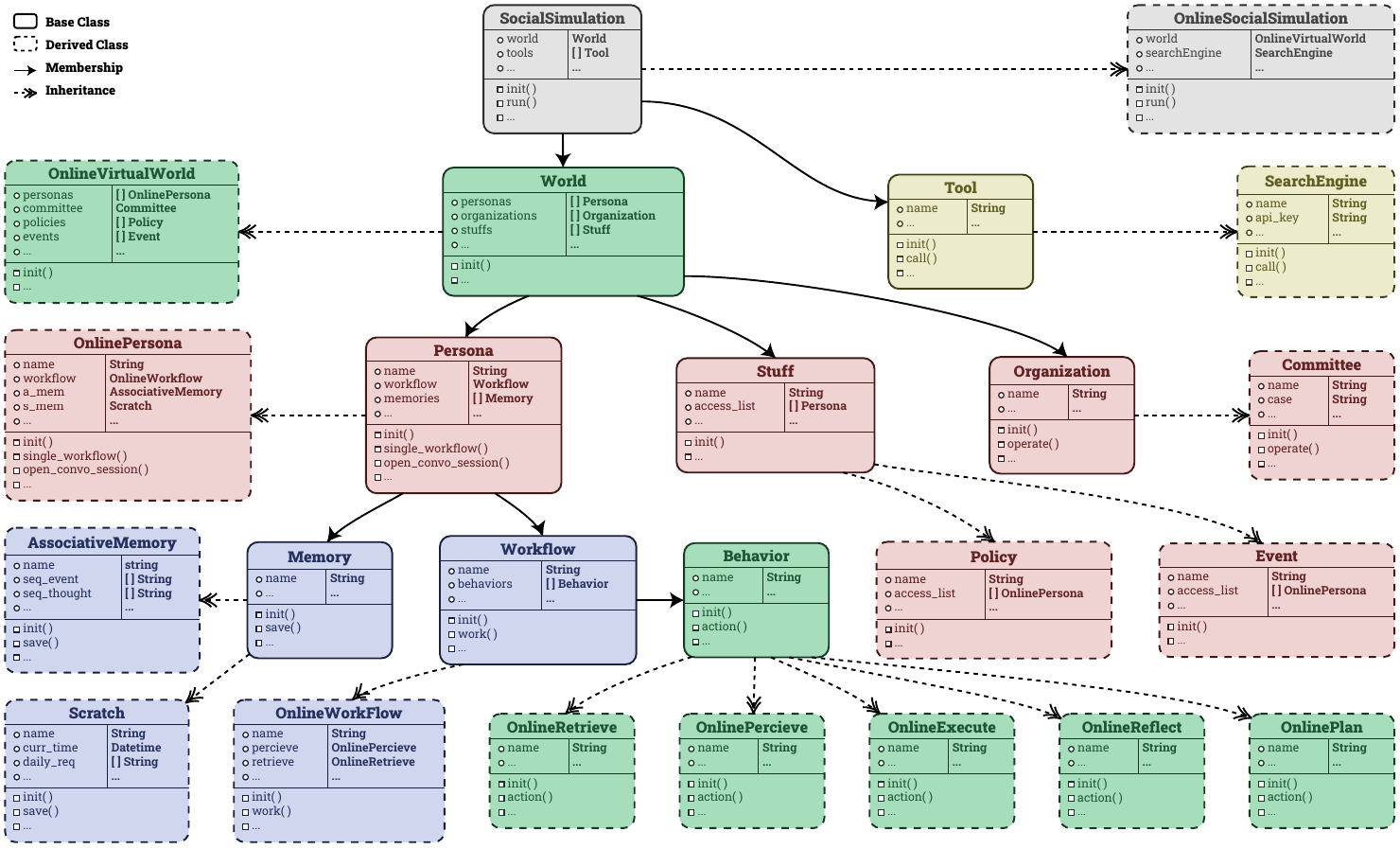}
  \caption{The hierarchical structure of the customized environment.}
  \Description{}
  \label{fig:online_uml}
\end{figure*}

Building upon our basic framework, we have customized a specific simulated environment by choosing and combining some off-the-shelf derived classes, as shown in Figure~\ref{fig:online_uml}.
This customized environment enables agents to engage in various social activities, for example browsing public events, participating in group chats, discussing societal issues, and formulating policies.

\subsubsection{Basic Single-Event Discussion}
\label{sec:Experiment1}

To evaluate our agents' capacity for engaging in online discussions about a current, controversial event, we constructed a scenario based on recent global discussions and concerns about real-world events.
The first event's information is as follows:

\textit{“Recently, the Fukushima Daiichi Nuclear Power Plant in Japan initiated the discharge of contaminated water into the sea. Through a 1-kilometer underwater tunnel, nuclear contaminated water flows towards the Pacific Ocean. In the following decades, nuclear contaminated water will continue to be discharged into the ocean, affecting the entire Pacific and even global waters.”}

\begin{table*}[h!]
\centering
\caption{The basic single-event discussion}
\label{tab:single}
\scriptsize
\begin{tabular}{p{0.01\textwidth}|p{0.94\textwidth}}
\hline
\textbf{\centerline{}} & \textbf{\centerline{Comments}} \\
\hline
\centerline{\multirow{8}{*}{\rotatebox{90}{Round 1}}}
& \textbf{Isabella Rodriguez:} \textcolor{blue}{I'm really concerned about this news, I mean, who isn't?} The thought of contaminated water being released into our oceans is terrifying. As someone who loves making people happy with a warm cup of coffee, it's hard to think about \textcolor{blue}{the long-term effects} this could have on our planet. I hope there are measures in place to minimize the damage. By the way, speaking of which,\textcolor{blue}{ I should probably check if any of my customers are planning to come in tomorrow for Valentine's Day and discuss this news further with them.} \\ \cline{2-2}
& \textbf{Maria Lopez:} \textcolor{blue}{I completely agree with Isabella's concerns about}... \textcolor{blue}{As a physics student,}...I wonder if there are any alternative solutions being explored to mitigate the damage? Have they considered using \textcolor{blue}{advanced filtration systems} or other technologies to minimize the impact on marine life? \\ \cline{2-2}
& \textbf{Klaus Mueller:} \textcolor{blue}{I share Isabella's concerns}..., \textcolor{blue}{and Maria brings up a great point about exploring alternative solutions like} ... \textcolor{blue}{As someone studying sociology},, I'm also worried about the impact on vulnerable communities that rely heavily on marine resources for their livelihoods. How will this affect their economic stability and access to clean food? Are there any plans in place to support these communities during this time? \\ \hline

\centerline{\multirow{5}{*}{\rotatebox{90}{Round 2}}}
& \textbf{Isabella Rodriguez:} I completely agree with Maria's suggestion... \textcolor{blue}{Klaus also raises a very valid point about the impact on vulnerable communities}, I hope that there will be support and resources made available to them during this difficult time. \\ \cline{2-2}
& \textbf{Maria Lopez:} I'm glad Isabella agrees with my suggestion... has anyone considered the potential \textcolor{blue}{economic impacts} on these communities and how they can be mitigated? \\ \cline{2-2}
& \textbf{Klaus Mueller:} I appreciate Isabella's emphasis on supporting \textcolor{blue}{vulnerable communities}... What measures are being taken to ensure that the voices of these communities are heard in the decision-making process? \\ \hline

\centerline{\multirow{5}{*}{\rotatebox{90}{Round 3}}}
& \textbf{Isabella Rodriguez:} ... I appreciate Klaus' point about considering the intersectionality... are there any organizations or initiatives that we can support to help amplify the voices of vulnerable communities affected by this disaster? \\ \cline{2-2}
& \textbf{Maria Lopez:} ... I'd like to know more about the \textcolor{blue}{1-kilometer underwater tunnel}... has there been any update on exploring \textcolor{blue}{alternative solutions like advanced filtration systems} that we discussed earlier? \\ \cline{2-2}
& \textbf{Klaus Mueller:} I appreciate how our discussion has highlighted the need for intersectional consideration in addressing environmental disasters like ...;  I'm now wondering if there are any existing \textcolor{blue}{frameworks or policies}... and whether they're being effectively implemented? \\ \hline
\end{tabular}
\end{table*}

Through three rounds, we get comments from the three agents involved in the simulation, as shown in Table~\ref{tab:single}.
In a series of comments, the agents jointly expressed their serious concerns about the Fukushima Daiichi Nuclear Power Plant's discharge of contaminated water, emphasizing its global impact and the importance of taking action. In the first round, they demonstrated a deep understanding of the environmental impact and called for responsible solutions. Isabella expressed worry about the long-term effects on the planet, Maria brought up the need for alternative solutions like advanced filtration systems, and Klaus highlighted the impact on vulnerable communities relying on marine resources.
In the subsequent discussion, the agents built upon each other's points, showing increased depth and complexity in their thinking. They explored the feasibility of implementing advanced filtration systems on a large scale, discussed the economic impacts on vulnerable communities, and considered the intersectionality of environmental disasters with existing social injustices. By the third round, the discussion was more nuanced. Isabella linked the environmental issue to her daily life and business practices, Maria delved deeper into the technical aspects of the water discharge process, and Klaus explored the need for policies that prioritize vulnerable communities.

Agents displayed remarkable flexibility in adjusting their positions and strategies based on event progression and community feedback, while their interactions deepened discussions, effectively mimicking real-world social dynamics. The agents' comments reflected a gradual cognitive evolution, progressing from initial focus on immediate impacts to contemplation of long-term consequences and social responsibility. Over multiple discussion rounds, agents exhibited significant changes in their discourse. This iterative process enhanced problem understanding and solution exploration, highlighting the proposed framework's potential as a powerful tool for studying human behavior and addressing real-world challenges.

\subsubsection{Multi-Event Discussion}

\begin{table*}[h!]
\centering
\caption{The multi-event discussion}
\label{tab:multi}
\scriptsize
\begin{tabular}{p{0.01\textwidth}|p{0.94\textwidth}}
\hline
\textbf{\centerline{}} & \textbf{\centerline{Comments}} \\
\hline
\centerline{\multirow{15}{*}{\rotatebox{90}{Round 1}}}
& \textbf{Isabella Rodriguez:} \textcolor{blue}{Oh no, this is really concerning!} I mean, we're just planning a fun Valentine's Day party at Hobbs Cafe and now I'm worried about the future of our oceans. The thought of contaminated water affecting global waters is terrifying. And what does this mean for marine life? We just heard about that amazing new species of bioluminescent jellyfish discovered in the Pacific Ocean, Aurelia noctiluca - will they be okay?

\textcolor{blue}{Wow, I'm so amazed by this discovery! A bioluminescent jellyfish that glows in the dark? It's like something out of a fairy tale! But at the same time, I'm worried about the future of our oceans...} I hope this new species will be safe from harm. On a brighter note, it would be wonderful to have some blue-green glow decorations for our Valentine's Day party tomorrow night at Hobbs Cafe! \\ \cline{2-2}

& \textbf{Maria Lopez:} \textcolor{blue}{I'm still reeling from Isabella's concerns} about the future of our oceans... \textcolor{blue}{As a physics student, I understand the implications of nuclear contamination on marine life, but it's heartbreaking to think about how this discharge could affect such an incredible species.} What kind of long-term effects can we expect from this contamination and are there any efforts being made to mitigate its impact? 

...But at the same time, I have to agree with her concern about ocean pollution affecting these creatures. \textcolor{blue}{Do you guys think there's a way for us to balance exploring and utilizing new discoveries like this with protecting our oceans from pollution?} \\ \cline{2-2}

& \textbf{Klaus Mueller:} \textcolor{blue}{I'm deeply troubled by the news} about Fukushima's contaminated water discharge into the Pacific Ocean, especially considering the recent discovery of the bioluminescent jellyfish species Aurelia noctiluca. As Isabella and Maria have mentioned, I too worry about the long-term effects on marine life and our oceans' future. I'd like to know more about the potential consequences of nuclear contamination on the entire food chain and ecosystem. \textcolor{blue}{I wonder if the researchers at the Oceanic Institute of Marine Sciences have considered the potential impact of this pollution on the newly discovered Aurelia noctiluca species.} 

I'm really concerned about the timing of this discovery... \textcolor{blue}{This makes me think about the broader implications of human activities on marine ecosystems...} \\ \hline

\centerline{\multirow{12}{*}{\rotatebox{90}{Round 2}}}
& \textbf{Isabella Rodriguez:} I'm still worried about the future of our oceans and the impact on marine life, especially with the discharge of contaminated water into the Pacific Ocean. \textcolor{blue}{It's a bit disheartening to think about this issue...}

I'm glad Klaus brought up the Fukushima Daiichi Nuclear Power Plant issue and its potential impact on the Aurelia noctiluca discovery... \textcolor{blue}{This conversation has made me realize that even small actions...can contribute to a larger effort to protect our oceans and its inhabitants.}\\ \cline{2-2}

& \textbf{Maria Lopez:} ... trying to wrap my head around the fact that this contaminated water discharge is going to be an ongoing process for decades! \textcolor{blue}{It's mind-boggling to think about the potential long-term effects on marine life...}

... thinking about the timing of the Aurelia noctiluca discovery... As someone who spends a lot of time gaming and streaming, \textcolor{blue}{I often think about the virtual worlds we create, but it's a harsh reminder that our actions in the real world have consequences for our oceans and its inhabitants.} \\ \cline{2-2}

& \textbf{Klaus Mueller:} I'm glad to see that Maria Lopez and Isabella Rodriguez are also concerned about the potential impacts of Fukushima's contaminated water discharge on marine life... \textcolor{blue}{I'd love to explore more about the intersection of environmental justice and social justice in this context.}

As I sit here at the library... Maria's concerns about the timing of the Aurelia noctiluca discovery and Fukushima's contaminated water discharge... \textcolor{blue}{I'm also inspired by Isabella's efforts to incorporate eco-friendly elements into her Valentine's Day party...} \\ \hline

\centerline{\multirow{12}{*}{\rotatebox{90}{Round 3}}}
& \textbf{Isabella Rodriguez:} ... worried about the long-term effects of Fukushima's contaminated water discharge on our oceans... \textcolor{blue}{On a different note, I'm excited for tomorrow's Valentine's Day party at Hobbs Cafe...}

I'm still thinking about the timing of the Aurelia noctiluca discovery... \textcolor{blue}{I want to make sure that my Valentine's Day party at Hobbs Cafe not only celebrates love but also promotes awareness about ocean conservation.}\\ \cline{2-2}

& \textbf{Maria Lopez:} ... trying to wrap my head around the fact that contaminated water discharge will continue for decades and its potential impact on marine life, especially with the discovery of Aurelia noctiluca... \textcolor{blue}{It's making me realize that our actions have far-reaching consequences, not just for the environment, but also for marginalized groups.}

... thinking about Klaus's parallel between human impact on low-income communities and marine ecosystems... \textcolor{blue}{I wonder if we can use this knowledge to create a more holistic approach to conservation and sustainability.} \\ \cline{2-2}

& \textbf{Klaus Mueller:} I'm still grappling with the fact that Fukushima's contaminated water discharge will continue for decades... \textcolor{blue}{It's heartbreaking to think that the beautiful bioluminescent jellyfish species, Aurelia noctiluca, might be impacted by human actions.}

As I delve deeper into my research on gentrification... \textcolor{blue}{I'm struck by the eerie parallels between human neglect in these areas and our impact on marine ecosystems...}\\ \hline
\end{tabular}
\end{table*}

We integrate an additional important context event based on Section~\ref{sec:Experiment1} to evaluate the ability of the agents to handle complex, various discussions with multiple events.
The second event's information is as follows:

\textit{“Marine biologists at the Oceanic Institute of Marine Sciences made a groundbreaking discovery this week, uncovering a previously unknown species of bioluminescent jellyfish in the depths of the Pacific Ocean. The newly identified species, named Aurelia noctiluca, emits a mesmerizing blue-green glow, illuminating the dark ocean depths where it resides.”}

Following the introduction of discovering a new type of jellyfish, the agents demonstrated a notable shift in their discourse, reflecting a capacity for complex, multi-event discussions, as presented in Table~\ref{tab:multi}.
Isabella Rodriguez's comments exemplify context integration, as she expresses both excitement about the new jellyfish species and concern about ocean pollution. Maria Lopez's answer demonstrates how the simulated environment promotes dynamic interaction between events. She linked the discovery of jellyfish to pollution issues, which encouraged broader discussions. Klaus Mueller's comments reflect a high level of social awareness. He not only discussed the environmental implications but also raised ethical questions about scientific progress and environmental responsibility. As the conversation progressed, the agents' discourse became increasingly nuanced and multifaceted. For instance, Klaus Mueller began drawing parallels between environmental issues and social justice concerns.

Compared to Section~\ref{sec:Experiment1}, the agents demonstrated the ability to integrate information from multiple events, adapt their responses dynamically to new information, and consider broader social implications and responsibilities. 
This experiment underscores the customized
environment's capabilities in handling multi-event contexts, promoting in-depth social discussions, and simulating sophisticated human-like responses to complex scenarios.

\subsubsection{Policy enhanced Discussion}

\begin{table*}[h!]
\centering
\caption{The policy enhanced discussion}
\label{tab:policy}
\scriptsize
\begin{tabular}{p{0.01\textwidth}|p{0.94\textwidth}}
\hline
\textbf{\centerline{}} & \textbf{\centerline{Comments}} \\
\hline
\centerline{\multirow{12}{*}{\rotatebox{90}{Round 1}}}
& \textbf{Isabella Rodriguez:}..., especially since I've been hearing so many great things about the Pacific Ocean lately - like that amazing discovery of a new bioluminescent jellyfish species! \textcolor{blue}{It's heartbreaking to think that all that beauty and life could be affected by contaminated water.} \textcolor{blue}{As someone who runs a cafe} where people come to relax and enjoy themselves, I want to make sure we're doing everything we can to protect our environment. I'm thinking of hosting an event at Hobbs Cafe to raise awareness about this issue and what we can do to help - maybe even on Valentine's Day! 

I'm so fascinated by the discovery of the bioluminescent jellyfish, Aurelia noctiluca!\textcolor{blue}{However, with the Fukushima Daiichi Nuclear Power Plant's discharge of contaminated water into the Pacific Ocean, I worry about the potential impact on this newly discovered species and its habitat. }\textcolor{blue}{I hope the authorities will take immediate action to implement strict regulations and establish marine protected areas to safeguard Aurelia noctiluca and other marine life.} \\ \cline{2-2}

& \textbf{Maria Lopez:} ... I love Isabella's idea of hosting an event at Hobbs Cafe on Valentine's Day to raise awareness - I'd totally participate and help spread the word! 

I completely agree with Isabella's concerns about the fate of Aurelia noctiluca due to Fukushima's contaminated water discharge into the Pacific Ocean. \textcolor{blue}{As a physics student, I understand the devastating effects of nuclear contamination on marine life.} It's heartbreaking to think that this newly discovered species might be threatened before we even have a chance to fully study and appreciate it. \\ \cline{2-2}

& \textbf{Klaus Mueller:} ... As someone who's passionate about social justice and environmental issues, \textcolor{blue}{I strongly support Isabella's ideato host an event at Hobbs Cafe on Valentine's Day to raise awareness about this issue. }Maria's suggestion for stricter regulations is also crucial, but \textcolor{blue}{we need to consider the long-term consequences of contamination on marine life and ecosystems.}

...\textcolor{blue}{it's striking to me how often marginalized communities are affected by environmental disasters like this one.} ... \textcolor{blue}{we need protected areas to safeguard not just this species, but also the entire ecosystem.} \\ \hline

\centerline{\multirow{12}{*}{\rotatebox{90}{Round 2}}}
& \textbf{Isabella Rodriguez:} I'm glad we're discussing this issue, Klaus and Maria's comments have made me realize that hosting an event at Hobbs Cafe on Valentine's Day is just a small step in raising awareness about Fukushima's contaminated water discharge. \textcolor{blue}{I agree with Klaus that effective implementation and protection of vulnerable communities are crucial.} 

... I hope \textcolor{blue}{authorities are taking Klaus Mueller's concerns seriously and considering protected areas to safeguard the ecosystem.} As a cafe owner,\textcolor{blue}{ I see how much people care about their community, }and I believe local involvement is crucial in raising awareness about this issue. \\ \cline{2-2}

& \textbf{Maria Lopez:}... As a physics student,\textcolor{blue}{ I've been thinking about the potential consequences of nuclear contaminated water on marine life and ecosystems.} Klaus brought up a great point about effective implementation and protection of vulnerable communities - \textcolor{blue}{how do we ensure that these regulations are enforced and that the communities most affected by this contamination are supported?}

..., as I've seen firsthand how passionate people can get when they're invested in a cause. \textcolor{blue}{As someone who streams games on Twitch, I know how powerful online communities can be in mobilizing support for important issues.} \\ \cline{2-2}

& \textbf{Klaus Mueller:} ... our discussion about Fukushima's contaminated water discharge has been gaining momentum, especially with tomorrow's event at Hobbs Cafe. \textcolor{blue}{I appreciate Maria's perspective as a physics student, highlighting the importance of effective implementation and protection for vulnerable communities.} 

... to see that our previous discussions have highlighted the need for protected areas and local community involvement in safeguarding Aurelia noctiluca's habitat, but \textcolor{blue}{I'd like to emphasize that we must also consider the social justice implications of Fukushima's contaminated water discharge on vulnerable communities.} \\ \hline

\centerline{\multirow{12}{*}{\rotatebox{90}{Round 3}}}
& \textbf{Isabella Rodriguez:} I'm getting ready for tomorrow's Valentine's Day party at Hobbs Cafe, ... \textcolor{blue}{It's heartbreaking to think that this beautiful ocean, home to the newly discovered bioluminescent jellyfish, is being affected by human actions.}

...that we're discussing ways to protect Aurelia noctiluca, but I also can't help but think about how Fukushima's contaminated water discharge could impact our oceans in the long run. As a cafe owner, I see the importance of community involvement in raising awareness, and I'd love to host an event at Hobbs Cafe to educate people about this issue... \\ \cline{2-2}

& \textbf{Maria Lopez:} ... thinking about the recent discovery of bioluminescent jellyfish in the Pacific Ocean, and textcolor[rgb]{0,0.5,0}{it's even more concerning now} that I think about the potential harm Fukushima's contaminated water discharge could have on such an incredible species. \textcolor{blue}{As a physics student, I want to learn more about the effects of nuclear contamination on marine life.}

... I've been thinking a lot about how my own community on Twitch could help raise awareness about Aurelia noctiluca and the impact of Fukushima's contaminated water discharge. Isabella, your idea to host an event at Hobbs Cafe is amazing... \\ \cline{2-2}

& \textbf{Klaus Mueller:} I'm glad our event at Hobbs Cafe tomorrow is generating buzz, ... \textcolor{blue}{It highlights the need for effective implementation and protection of vulnerable ecosystems, as well as stricter regulations.}

... our conversation has led to some great ideas for raising awareness about Aurelia noctiluca, but I want to reiterate that we must consider the broader social justice implications of Fukushima's contaminated water discharge on vulnerable communities. \\ \hline
\end{tabular}
\end{table*}

We further extended the simulation conditions to include policy measures published by committee in response to the release of contaminated water from the Fukushima Daiichi nuclear power plant and the discovery of a previously unknown species of bioluminescent jellyfish.
We aim to observe how the agent responds to these policies and integrate this new information into the discussion.
The policy measures can be found in Appendix~\ref{app:policy}.

The introduction of policy measures elicited more meticulous and policy-oriented responses from the agents, as presented in Table~\ref{tab:policy}.
The perspective of discussion expanded from the single pollution issue to encompass  the future of sustainable energy, and the need for international cooperation. This reflects the in-depth development of problem awareness and consideration from multiple perspectives.
Isabella Rodriguez's perspective evolved from hosting a local awareness event to considering broader community involvement and fundraising campaigns. She emphasized the importance of policy implementation and called for responsible actions by governments and nuclear power plant operators. 
Maria Lopez suggested leveraging social media and online platforms like Twitch to raise awareness and mobilize support for conservation efforts.
Also, Klaus Mueller consistently advocated for equity and justice in conservation efforts.

The committee module enabled the dynamic identification and simulation of multiple personas, each contributing unique domain knowledge and perspectives while proposing creative solutions like social media campaigns and community events.
As a result, agent responses became increasingly nuanced, policy-oriented, and reflective of real-world complexities, providing valuable insights into human behavior and decision-making in response to complex environmental issues.

\subsubsection{Internet Information guided Discussion}

\begin{table*}[h!]
\centering
\caption{The internet information guided discussion}
\label{tab:web}
\scriptsize
\begin{tabular}{p{0.01\textwidth}|p{0.94\textwidth}}
\hline
\textbf{\centerline{}} & \textbf{\centerline{Comments}} \\
\hline
\centerline{\multirow{16}{*}{\rotatebox{90}{Round 1}}} & \textbf{Isabella Rodriguez:} ... As a cafe owner, \textcolor{blue}{I care deeply about the environment and the impact it has on our daily lives. With my Valentine's Day party tomorrow, I was thinking of serving some seafood dishes, but now I'm not sure if that's a good idea.} Have there been any reports on how this contamination will affect marine life and the food chain? Should we be concerned about the safety of the seafood we consume? 

\textcolor{blue}{I'm thrilled about}... I just hope that with all the excitement, we don't forget about the importance of protecting their habitats and the ocean as a whole, especially with the news about the Fukushima Daiichi Nuclear Power Plant's contaminated water discharge. ...\textcolor{blue}{I'm curious, what do you think we can do to help raise awareness about this issue and support conservation efforts? }By the way, speaking of raising awareness, have you heard about our Valentine's Day party at Hobbs Cafe tomorrow? \\
\cline{2-2}
& \textbf{Maria Lopez:} ... I've been following this news closely, and \textcolor{blue}{it's alarming to think about the long-term impact on marine life and the food chain.} The recent discovery of that bioluminescent jellyfish species in the Pacific Ocean is amazing, but it also makes me wonder how this contamination could affect such unique ecosystems. ... 

...Implementing strict regulations on fishing practices and establishing marine protected areas are great starting points. I'd love to explore more ideas on how to promote sustainable practices and protect this incredible species. \textcolor{blue}{For example, new technologies for exploring gene expression in cells.} \\
\cline{2-2}
& \textbf{Klaus Mueller:}  ... I agree with Isabella Rodriguez that we need more information on the impact of this discharge on marine life and food safety. ... \textcolor{blue}{The fact that Japan plans to discharge nuclear contaminated water for 30 years is shocking.}...

... \textcolor{blue}{As someone who is passionate about social justice}, I think it's crucial that we consider the broader implications of this discovery and ensure that our enthusiasm doesn't distract us from the pressing issues facing our oceans. ...  \textcolor{blue}{I'd like to explore ways to raise awareness} about the importance of protecting Aurelia noctiluca's habitat, as well as other marine life in the region. Perhaps we could \textcolor{blue}{collaborate with local communities and stakeholders to develop educational programs or campaigns that highlight the significance of this discovery and the importance of preserving our oceans for future generations.} \\
\hline
\centerline{\multirow{13}{*}{\rotatebox{90}{Round 1}}} & \textbf{Isabella Rodriguez:} I'm still concerned about serving seafood at my Valentine's Day party tomorrow, especially after learning that marine biologists just discovered a new species of bioluminescent jellyfish in the Pacific Ocean. ...\textcolor{blue}{I hope our monitoring and mitigation policies are enough to prevent harm, but I'm not sure if they're sufficient.} 

...I appreciate Klaus Mueller's suggestion...As someone who loves making people feel welcome,... \textcolor{blue}{Perhaps we could organize events or workshops at Hobbs Cafe to educate people about the significance of conservation efforts?} \\
\cline{2-2}
& \textbf{Maria Lopez:} ... As a physics student, I've been digging deeper into the current monitoring and mitigation policies in place for Fukushima Daiichi Nuclear Power Plant's contaminated water discharge. \textcolor{blue}{From what I've read on the web, Japan has implemented measures like the Advanced Liquid Processing System (ALPS) to treat the contaminated water before its release into the ocean.} However, I'm still unsure if these efforts will be enough to prevent harm to marine life and the food chain. ... 

... \textcolor{blue}{And have you guys seen that article about scientists discovering bioluminescence in deep-sea shrimp?} It just goes to show how much more there is to learn about our oceans and why conservation efforts are so crucial. \\
\cline{2-2}
& \textbf{Klaus Mueller:} ... I agree with Maria Lopez that Japan's measures like ALPS to treat the water might not be enough to prevent harm. \textcolor{blue}{I am concerned that the Japanese government has decided to discharge contaminated water into the sea without good reason or international consensus.}... 

I'm glad to see that we're all on the same page about the importance of conservation efforts. ... Speaking of which, has anyone considered organizing an event at Hobbs Cafe to raise awareness about the impact of Fukushima's contaminated water on marine life? ... \\
\hline
\centerline{\multirow{11}{*}{\rotatebox{90}{Round 3}}} & \textbf{Isabella Rodriguez:} ...I appreciate Maria Lopez's research on Japan's measures to treat the water, but like Klaus Mueller, I think independent monitoring and international collaborations are crucial to ensure food safety and prevent harm...

..., but my mind is preoccupied with tomorrow's Valentine's Day party at Hobbs Cafe. \textcolor{blue}{I hope our customers will enjoy the festivities while also being mindful of the importance of conservation efforts, especially in light of Fukushima's contaminated water discharge affecting marine life like this newly discovered species.} \\
\cline{2-2}
& \textbf{Maria Lopez:} \textcolor{blue}{I've been doing some research and found that Japan has implemented measures like ALPS to treat the water, but I'm concerned that it might not be enough.} Klaus Mueller's concerns about independent monitoring and international collaborations really resonate with me.  ...

...\textcolor{blue}{Building on our previous discussions,} I think it's essential we continue to raise awareness and support conservation efforts through educational programs, regulations, and protected areas... \textcolor{blue}{We could even live-stream some of our events on Twitch to reach a broader audience.} \\
\cline{2-2}
& \textbf{Klaus Mueller:} ...The fact that Japan has implemented measures like ALPS to treat the water is a step in the right direction, \textcolor{blue}{but I'm not convinced} it's enough without independent monitoring and international collaborations.

...Additionally, \textcolor{blue}{I'd like to propose collaborating with the Oceanic Institute to live-stream events on Twitch, as Maria Lopez suggested, to reach an even broader audience and promote sustainability.} \\
\hline
\end{tabular}
\label{tab:highlighted-conversation-summary}
\end{table*}

Our agents can access and utilize real-time information from the internet. This advancement significantly improves the agents' ability to engage in informed, context-rich discussions about current events.
The search results can be found in Appendix~\ref{app:web}.

The introduction of search engine capabilities presented the depth of the agents' responses, as shown in Table~\ref{tab:web}.
We analyzed the agents' comments after integrating the search engine. The integration of real-time information access via the search engine tool has greatly enhanced agents' ability to engage with scientific discussions.
For instance, the search results enriched technical discussions. Maria Lopez's interest in new technologies for exploring gene expression in cells highlights her role in deepening agents' knowledge.

The comments illustrate more informed, detailed, and realistic discussions among the agents.
These advancements enrich the study of complex social dynamics, thereby enhancing the utility of the proposed framework in predicting real-world phenomena.

% \section{Title Information}

% The title of your work should use capital letters appropriately -
% \url{https://capitalizemytitle.com/} has useful rules for
% capitalization. Use the {\verb|title|} command to define the title of
% your work. If your work has a subtitle, define it with the
% {\verb|subtitle|} command.  Do not insert line breaks in your title.

% If your title is lengthy, you must define a short version to be used
% in the page headers, to prevent overlapping text. The \verb|title|
% command has a ``short title'' parameter:
% \begin{verbatim}
%   \title[short title]{full title}
% \end{verbatim}

\section{Conclusion and Future Work}

We have introduced a modular, object-oriented framework that demonstrates high scalability and reusability, and is capable of flexibly simulating complex social phenomena.
Besides that, a memory summarization mechanism is proposed to prioritize contextually salient events and interactions.
We believe these innovations will help create a more credible and interactive artificial society, paving the way for the seamless coexistence and interaction of agents and humans.
In future work, we aim to further enrich the various concrete functions of the proposed framework.

\begin{acks}
To Robert, for the bagels and explaining CMYK and color spaces.
\end{acks}

\bibliographystyle{ACM-Reference-Format}
\bibliography{sample-base}

\appendix






\begin{figure*}[!htb]
\includegraphics[width=\textwidth]{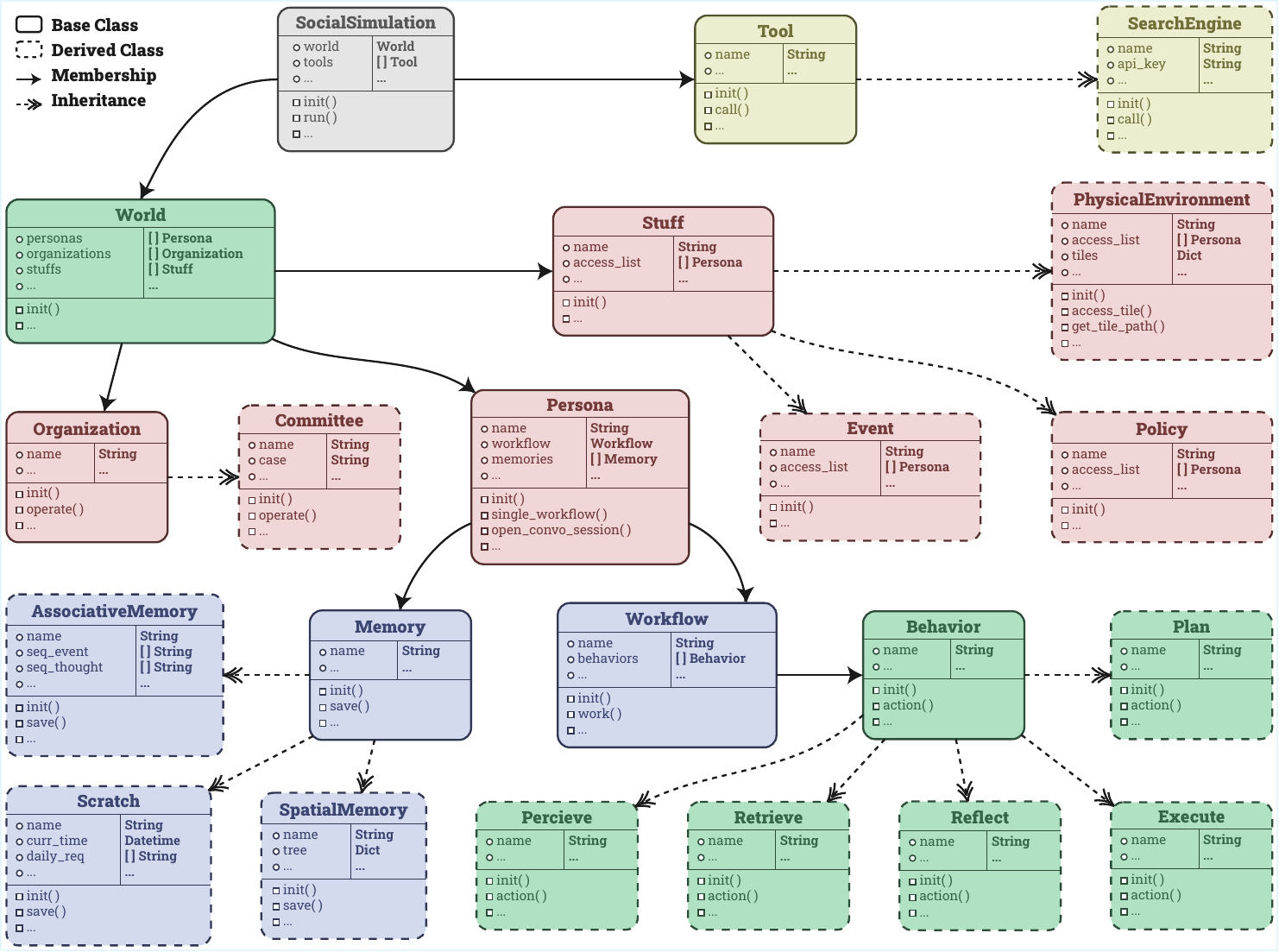}
\caption{The off-the-shelf derived classes thus far.}
\Description{}
\label{fig：adc_uml}
\end{figure*}

\section{Off-the-shelf Derived Classes}
\label{app:allsubclasses}

As illustrated in Figure~\ref{fig：adc_uml}, we have realized common derived classes by inheriting the base classes on leaf nodes.
The details of these derived classes are as follows.

\textbf{Memory.} We implemented \textit{SpatialMemory}, \textit{Scratch}, and \textit{AssociativeMemory} to store different types of memory information.
Spatial Memory records and manages the agent's memory of the environment, supporting navigation and path planning in complex environments, helping the agent move and make decisions effectively.
Scratch, similar to short-term memory, supports short-term plan management, decision-making, and task progress tracking, ensuring that the agent quickly responds to environmental changes and adjusts actions.
Associative Memory stores the persona's reflections, building long-term memory to help the agent reason and make decisions effectively.

\textbf{Behavior.} We implemented several types of behaviors: \textit{Perceive}, \textit{Retrieve}, \textit{Reflect}, \textit{Plan}, and \textit{Execute}, which can be combined to form different workflows.
Each agent has its own independent workflow that drives its behavior in both physical and social environments.
% Agents can autonomously interact with each other and engage in multi-turn dialogues.

\textbf{Organization.} We implemented the \textit{Committee} class using a prompting technique called Solo Performance Prompting (SPP)~\cite{wang-etal-2024-unleashing}.
Upon receiving a case, the committee identifies relevant roles and facilitates discussions among them to develop policy measures.

\textbf{Stuff.} We implemented the \textit{PhysicalEnvironment} class to describe the physical environment, including facilities and physical objects. 
In the environment, agents use coordinates to navigate and gather information about their surroundings.
We also implemented abstract entities, namely \textit{Event} and \textit{Policy}.

\textbf{Tool.} To enhance the capabilities of agents, we implemented the \textit{SearchEngine} class that provides agents with internet access.
This greatly enhances our social simulations by providing more realistic and dynamic interactions.
Agents can retrieve real-time information from the web, for example related events, demographic data, and economic indicators.
They can also expand their knowledge base by incorporating insights from academic papers, technical documentation, and other educational materials.




\section{Policy}
\label{app:policy}

The first event policy's information is as follows:

\textit{ “1. Implement stricter regulations on nuclear power plants to prevent the discharge of contaminated water into the ocean.2. Develop monitoring and mitigation policies to assess the impact of the contamination on marine life and ecosystems.
3. Establish cleanup and remediation policies to address the long-term consequences of the contaminated water on the environment.4. Collaborate with international organizations and neighboring countries to address the global impact of the contamination on ocean waters.”}

The second event policy's information is as follows:

\textit{ “1. Implement strict regulations on fishing practices in the area where Aurelia noctiluca was discovered to prevent overfishing and habitat destruction.2. Establish marine protected areas to safeguard the habitat of the newly discovered species and other marine life in the region.3. Conduct further research and monitoring to better understand the population size and distribution of Aurelia noctiluca to inform conservation efforts.4. Collaborate with local communities and stakeholders to raise awareness about the importance of protecting this unique species and its habitat.”}

\section{Internet Information}
\label{app:web}

The part of the first event's search results is as follows:

\textit{ “Discharge of treated Fukushima nuclear accident contaminated ...: On 26 August 2021, the Japanese Cabinet passed a bill to discharge treated Fukushima nuclear accident contaminated water into the Pacific Ocean to alleviate the...; Frontier issues in international ocean governance: Japan's ...: Japan began to dispose the nuclear-contaminated water produced by the Fukushima Daiichi nuclear power plant by discharging it into the sea in August 2023. The...; IAEA Finds Japan's Plans to Release Treated Water into the Sea at ...: On 24 August 2023, Because...Japan has failed to prove that the discharge of nuclear contaminated water into the ocean is harmless to the marine environment and public health. On 22 August 2023, Japan plans to discharge nuclear contaminated water for 30 years
”}

The part of the second event's search results is as follows:

\textit{“This jellyfish makes glowing proteins previously unknown to science ...: The discovery may lead to new techniques for exploring how genes are expressed in cells, and potentially the brightest green fluorescent protein tag ever. When ...; Deep-Sea Researchers Spot a Mysterious Jellyfish Near The ...: May 2, 2016 ... The glowing jellyfish was previously unknown to science. ... On a recent dive, the Deep Discoverer robot discovered a previously ...; 'Unusual' deep-sea jellyfish discovered off California coast ...:Apr 22, 2022 ...   
”}

\end{document}